**Subject areas:**

Physical sciences (Optics and photonics / Nanoscience and technology)

# Randomness in highly reflective silver nanoparticles and their localized optical fields

Makoto Naruse[1,2,a)], Takeharu Tani[3], Hideki Yasuda[3], Naoya Tate[2,4], Motoichi Ohtsu[2,4], and Masayuki Naya[3]

1 Photonic Network Research Institute, National Institute of Information and Communications Technology, 4-2-1 Nukui-kita, Koganei, Tokyo 184-8795, Japan

2 Nanophotonics Research Center, Graduate School of Engineering, The University of Tokyo, 2-11-16 Yayoi, Bunkyo-ku, Tokyo 113-8656, Japan

3 Frontier Core-Technology Laboratories, Research and Development Management Headquarters, Fujifilm Corporation, Nakanuma, Minamiashigara, Kanagawa 250-0193, Japan

4 Department of Electrical Engineering and Information Systems, Graduate School of Engineering, The University of Tokyo, 2-11-16 Yayoi, Bunkyo-ku, Tokyo 113-8656, Japan

a) Electronic mail: naruse@nict.go.jp

**Abstract: Reflection of near-infrared light is important for preventing heat transfer in energy saving applications. A large-area, mass-producible reflector that contains randomly distributed disk-shaped silver nanoparticles and that exhibits high reflection at near-infrared wavelengths was demonstrated. Although resonant coupling between incident light and the nanostructure of the reflector plays some role, what is more important is the geometrical randomness of the nanoparticles, which serves as the origin of a particle-dependent localization and hierarchical distribution of optical near-fields in the vicinity of the nanostructure. Here we show and clarified the unique optical near-field processes associated with the randomness seen in experimentally fabricated silver nanostructures by adapting a rigorous theory of optical near-fields based on an angular spectrum and detailed electromagnetic calculations.**

Near-infrared light-reflecting films attached to windows etc. are important for preventing heat transfer from sunlight, thus saving energy for cooling rooms in summer[1-4]. At the same time, it is important to maintain higher transmission efficiency in the visible range to ensure good visibility, as well as in radio frequency bands so as not to interfere with wireless communications. A large area and mass-producibility are also critical for market deployment. Fujifilm Co. Ltd. has proposed and realized a device named NASIP (Nano Silver Pavement), which satisfies all of the above requirements, consisting of randomly distributed disk-shaped silver nanoparticles for reflecting near-infrared light[1]. An actual device is shown in Fig. 1a, and a scanning electron microscope image of the surface is shown in Fig. 1b[2]. The elemental silver nanostructure has a diameter of about 120 to 150 nm and a thickness of 10 nm. From Fig. 1b, we can see that the nanoparticles contain randomness in terms of their individual shapes and layout. Figure 1c shows experimentally observed spectral properties of the fabricated device, which exhibits high reflectance for near-infrared light while maintaining a high transmittance in the visible and far-infrared regions.

The physical principle of the device has been attributed to plasmon resonances between the incoming light and the individual nanostructured matter[1]. However, as introduced later with Fig. 2b, whereas strong light localization indeed occurs in certain nanoparticles at the resonant wavelengths, not-so-evident localization has also been observed in some nanoparticles. Also, strong localization is observed even at non-resonant wavelengths in some nanoparticles. Such features may be attributed to the above-mentioned randomness of nanostructures. Furthermore, unlike uniformly distributed, uniformly shaped nanoparticles, interesting optical near-field distributions are present in a device containing geometrical randomness.

In this paper, by investigating the inherent randomness of silver nanoparticles while adapting a rigorous theory of optical near-fields based on an angular spectrum, we clarified the unique optical near-field processes associated with randomness in nanostructures, which are not observed in uniformly arranged nanostructures. Unlike the well-known Anderson localization[5] scheme, where multiple coherent scattering and the constructive interference of certain scattering paths are the origin of the localization phenomena[6], with our approach we are able to highlight near-field interactions and "hierarchical" attributes in the vicinity of nanostructures. The word "hierarchical" here indicates that localized electromagnetic fields exist in a plane distant from the surface of the nanostructure; such a feature is manifested via the angular-spectrum-based theory, as shown below. Regarding random media and light localization, Grésillon *et al.* did pioneering work by evaluating a metal–dielectric film having a random surface profile using near-field scanning microscopy[7,8]. Birowosuto *et al.* observed fluctuations in the local density of states in random photonic media[9], and Krachmalnicoff did so on disordered metal films[10]. With respect to statistical insights in related fields, Le Ru *et al.* investigated a power law distribution of electromagnetic enhancement and its relation to the detection of single molecules[11], and Sapienza *et al.* investigated the long-tail statistics of the Purcell factor in disordered media[12]. With our work, we would like to contribute to the understanding of the fundamental aspects of randomness in an array of nanoparticles and their localized optical fields by adapting a rigorous theory of optical near-fields based on an angular spectrum and detailed electromagnetic calculations.

It is well-known that the resonance between far-field light and metal nanostructures depends on the type of metal, geometries such as size or shape, environmental materials, and so forth[13]. Silver has been chosen for such devices since it exhibits a remarkable resonance with near-infrared light[1]. Regarding planar metal nanostructures, it is known that the resonant wavelength can be engineered in the range from visible to infrared by modifying the aspect ratio [14]. A 10 nm-thick layer containing about 100 nm-diameter silver nanoparticles, shown in Fig. 1b, which corresponds to an aspect ratio of 10, realizes a resonant wavelength in the near-infrared region. We assume a surrounding dielectric material whose refractive index is 1.5.

In order to characterize the electromagnetic properties associated with experimentally fabricated devices in detail, the geometries of the fabricated silver nanoparticles are converted to a numerical model consisting of a vast number of voxels. Specifically, the SEM image shown in Fig. 1b, which occupies an area of 4.2 μm × 4.2 μm, is digitized into binary values with a spatial resolution of 2.5 nm both horizontally and vertically. The pixels occupied by silver take values of one, whereas those specified by the substrate material take values of zero. The thickness of the silver nanoparticles may exhibit certain position-dependent variations, as indicated by the grayscale differences observed in Fig. 1b; but we consider that they do not significantly affect the overall optical properties, and thus assume that the thickness is constant.

As a result, the silver nanoparticles, of which total number $N$ is 468, are numerically modeled by 4200 × 4200 × 5 voxels in an *xyz* Cartesian coordinate system, giving a total of 88.2 M voxels. This model is then simulated with a finite-difference time-domain-based electromagnetic simulator with assuming continuous-wave (CW) *x*-polarized light incident normally on the surface of the silver nanostructures (Fig. 2a). The details are shown in Methods section. Figure 2b summarizes electromagnetic intensity distributions calculated at a position 5 nm away from the surface of the silver nanoparticles on which the input light is incident. The wavelength is from 300 nm to 2000 nm in 100 nm intervals. Around the wavelength of 1000 nm, highly localized electric fields are observed in the vicinity of silver nanoparticles, but it should also be noted that not all nanoparticles have accompanying high-intensity fields. Also, even at some off-resonant wavelengths, most of the nanoparticles carry low-intensity electric fields, but a few of them carry high-intensity electric fields.

In order to characterize the localization of optical near-fields stemming from the geometrical randomness of the silver nanoparticles, we take the following strategy. First, we derive the induced charge distributions in silver nanostructures. Specifically, we calculate the divergence of the calculated electric fields at planes within the silver nanostructure, and sum up along the Z-direction to represent the charge density distributions. Figure 3a shows such a charge distribution, denoted by $\rho(x, y)$ brought about by 1000 nm-wavelength light.

Now, we characterize the induced charge distribution at each nanoparticle as an oscillating electric dipole. Figure 3b schematically illustrates a nanoparticle identified by index $i$ ($i = 1, \cdots, N$). The position of the center of gravity of the nanoparticle $i$ is given by

$$G^{(i)} = \left( \frac{\sum_{P^{(i)}} xp(x,y)}{\sum_{P^{(i)}} p(x,y)}, \frac{\sum_{P^{(i)}} yp(x,y)}{\sum_{P^{(i)}} p(x,y)} \right), \quad p(x,y) = \begin{cases} 1 & (x,y) \in P^{(i)} \\ 0 & \text{otherwise} \end{cases} \quad (1)$$

where $P^{(i)}$ indicates the area that the nanoparticle $i$ occupies. We consider that the imbalance of the induced charge regarding the vertical and horizontal half spaces divided by $G^{(i)}$ is represented by an induced dipole given by

$$\boldsymbol{d}^{(i)} = \left( \sum_{x \geq G^{(i)}(x)} \rho(x,y) - \sum_{x < G^{(i)}(x)} \rho(x,y), \sum_{y \geq G^{(i)}(y)} \rho(x,y) - \sum_{y < G^{(i)}(y)} \rho(x,y) \right) = d^{(i)} \exp(i\phi^{(i)}) \quad (2)$$

where $d^{(i)}$ and $\phi^{(i)}$ ($-\pi < \phi^{(i)} \leq \pi$) respectively represent the magnitude and phase of $\boldsymbol{d}^{(i)}$. Here the charge density distributions, $\rho(x,y)$, are temporally oscillating variables; we represent $d^{(i)}$ by its maximum value during a single period of lightwave oscillation, and $\phi^{(i)}$ is determined accordingly. Figure 3c schematically indicates the calculated dipoles, where both the magnitude and the phase seems to contain certain randomness.

To quantify such randomness, the incidence patterns of the magnitude and the phase of the induced dipoles are analyzed as shown respectively in Fig. 4a and Fig. 4b with respect to all wavelengths. (The incidence patterns at the wavelengths of 300 nm, 500 nm, ..., and 1900 nm are not shown.) The magnitude of the dipoles exhibits some variation around the wavelength of 1200 nm, and the phase shows significant variation at 1000 nm. The square and circular marks in Fig. 4c and Fig. 4d respectively represent the average and the standard deviation of the magnitude and the phase of the dipoles. From Fig. 4d, the average phase is changed by $\pi$ for 1000 nm light, which might be a manifestation of the resonance between the incident light and the silver nanostructured matter. If the nanoparticles have the same shape and are arranged uniformly, no such diversity in induced dipoles is observed; therefore, it is concluded that the geometrical randomness of the silver nanostructure must contribute to these electromagnetic characteristics.

Moreover, such variations in phase or magnitude, or both, lead to a variety of electronic localizations of the electronic fields, not just in the close vicinity of the silver nanostructures shown in Fig. 2b. We also evaluate the electric field intensity distributions at planes distant from the surface by 20, 50, 100, 200, 500, and 1000 nm, respectively, for all wavelengths. For these analyses, we conducted other numerical simulations by converting the above 2.5 nm-grid silver nanostructure model to a 10 nm-grid model. In addition, 1100 nm-thick and 1090 nm-thick volumes were assumed in the electromagnetic calculation above and beneath the silver nanostructures. The CW light source was located 1070 nm away from the surface of the silver nanostructure.

Figure 5 summarizes incidence patterns of the electric field intensity at different heights for all wavelengths. (The patterns with respect to the wavelengths of 300 nm, 500 nm, ... , and 1900 nm are not shown.) It is evident that the electric field intensity exhibits a variety of values even 100 nm away, for the wavelengths between 1000 nm and 2000 nm. As the distance from the surface increases to larger than 200 nm, the variation in the electric field intensity decreases. However, it should be noted that at the resonant frequency, at the wavelength of 1000 nm, even a distant plane (1000 nm away from the silver nanoparticles) contains some variation in the electric field intensity. Such a nature has never been found in the electric fields accompanying uniform-shaped, uniformly distributed nanostructures. In the following, we provide the theoretical background behind such attributes.

The angular spectrum representation of an electromagnetic field involves decomposing an optical field as a superposition of plane waves including evanescent components[15,16], which allows us to explicitly represent and quantify optical near-fields. The details of the angular spectrum representation used in this paper are shown in the Supplementary Information and in Fig. 6a. The electric field $E(r)$ originated by a point dipole $d^{(i)}$ with frequency $K$ is given in the form[10]

$$E(r) = \left(\frac{iK^3}{8\pi^2\varepsilon_0}\right) \sum_{\mu=TE}^{TM} \iint_{-\infty}^{-\infty} ds_x ds_y \frac{1}{s_z} \left[\varepsilon(s^{(+)},\mu) \cdot d^{(i)}\right] \varepsilon(s^{(+)},\mu) \exp(iKs^{(+)} \cdot r). \quad (3)$$

Suppose that the dipole is oriented in the $xz$ plane and is given by $d^{(i)} = d^{(i)}(\sin\theta^{(i)}, 0, \cos\theta^{(i)})$, and that the point $r$ is also on the $xz$ plane and is given by the displacement from the dipole, or $R^{(i)} = (X^{(i)}, 0, Z^{(i)})$ (Fig. 6b). In such a case, the angular spectrum representation of the $z$ component of the electric field in the evanescent regime (namely, $1 \leq s_\parallel < +\infty$) based on eq. (3) is given by[17,18]

$$E_z(R^{(i)}) = \left(\frac{iK^3}{4\pi\varepsilon_0}\right) \int_1^\infty ds_\parallel \frac{s_\parallel}{s_z} f_z^{(i)}(s_\parallel, d^{(i)}), \quad (4)$$

where

$$f_z^{(i)}(s_\parallel, d^{(i)}) = ds_\parallel \sqrt{s_\parallel^2 - 1} \sin\theta^{(i)} J_1\left(KX^{(i)}s_\parallel\right) \exp\left(-KZ^{(i)}\sqrt{s_\parallel^2-1}\right) + ds_\parallel^2 \cos\theta^{(i)} J_0\left(KX^{(i)}s_\parallel\right) \exp\left(-KZ^{(i)}\sqrt{s_\parallel^2-1}\right). \quad (5)$$

Here, $J_n(x)$ represents a Bessel function of the first kind, where $n$ is an integer. One minor remark here is that the dipole model derived in eq. (2) spans in the $xy$ plane, whereas the dipole used in the above angular spectrum theory is oriented in the $xz$ plane. Although it is possible to assume a dipole oriented in the $xy$ plane in the angular spectrum theory[16,17], we consider that the resulting mathematical representations and the formula for the angular spectrum would be unnecessarily much more complex, whereas assuming dipoles in the *xz* plane preserves the essential attributes of the discussion of randomness in this paper while keeping the mathematical formula simpler.

In order to examine the character of optical near-fields that originate from the structural randomness, we consider two virtual dipoles $d^{(1)}$ and $d^{(2)}$ located on the *x*-axis and separated from each other by a small distance *G*, as shown in Fig. 6c. Now, supposing that the magnitude and the orientation of these dipoles are chosen in a random manner, we investigate the resultant near-fields. Specifically, we evaluate the angular spectra at a position equidistant from both $d^{(1)}$ and $d^{(2)}$ and a distance Z away from the *x*-axis. Let the magnitude of a dipole be given by a random number that follows a normal distribution whose average and standard deviation are given by 2 and 1/2, respectively. The phase is also specified by a random number following a normal distribution whose average and standard deviation are both given by π/2. We characterized 10,000 kinds of such angular spectra based on pairs of dipoles, each of which was randomly chosen. Figure 7a shows ten example angular spectra obtained when *G* was λ/4 and Z was λ/16, whose differences are evident.

On the other hand, when the randomness associated with the dipoles is smaller, for example, in the case of the amplitude, the random number follows a normal distribution with an average of 2 and a standard deviation of 1/20, and in the case of the phase, the random number follows a normal distribution with average π/2 and standard deviation π/20. Ten example angular spectra, out of 10,000, are shown in Fig. 7b, where the differences among the angular spectra are considerably smaller.

The integral of the angular spectrum along the spatial frequency is correlated with the electric field intensity[16]. Figures 7c,i and 7d,i respectively represent the incidence patterns of the integral of 10,000 kinds of angular spectra, showing the variation in electric field intensity, with respect to the former highly random dipole pairs and the latter less-random ones. It should be noted that considerable variation appears in the case of the higher randomness, whereas the electric field intensity is almost uniquely low with the less-random dipoles. As the distance from the dipole, namely Z, increases, the variation of the angular spectra, and their resultant electric field intensity, decreases. However, as is shown in Figs. 7c,ii and 7c,iii, which are for the cases $Z=\lambda/8$ and $Z=\lambda/2$, respectively, the electric field intensity histogram still contains some variation; that is to say, highly localized optical near-fields could exist in some cases. On the other hand, such an intensity variation is *never* observed in the case of less-randomly formed dipole pairs, as shown in Figs. 7d,ii and 7d,iii. These theoretically obtained characteristics agree well with the former results shown in Fig. 5 derived by electromagnetic calculations based on experimentally observed randomly organized silver nanostructures.

One minor remark that we should make regarding the analysis is that the distance between the two dipoles is kept constant, and this distance could be set in a random manner. We consider that the geometrical randomness causes the imbalance of the magnitude and the phase of dipoles, and thus, investigating the attributes of optical near-fields by taking account of the randomness associated with dipoles is the most important matter at this stage. The geometrical randomness of the nanostructures and the randomness in the induced dipoles may have certain complex correlations, and such issues should be investigated in future work.

Finally, the diamond and triangular marks in Fig. 8 show the reflection efficiencies based on electromagnetic calculations in the cases of the random structure based on the experimentally observed silver nanoparticles and an "ordered" structure, respectively. Here, the "ordered" structure consisted of an array of constant-diameter (128 nm), 10 nm-thick silver nanoparticles, and the total area and the total number of nanostructures were equivalent to the case of the random structure. Although the reflectance of the ordered structure exhibits a higher value than that of the random structure at the resonant frequency (1000 nm), the average reflectance between the wavelengths of 1000 nm and 2000 nm results in a higher value in the case of the random structure (0.171) than in the case of the ordered one (0.138). The square and circular marks in Fig. 8 represent the transmittances for the random and ordered structures, respectively, whose averages in the spectral range between 1000 nm and 2000 nm are given by 0.710 and 0.805, meaning that the random structure provides lower transmittance. Thus, we have shown that a silver nanostructured material having randomness yields different performance figures of the near-infrared reflection film compared with those of a reflection film containing an equivalent amount of uniformly distributed silver nanoparticles.

In summary, we examined the optical near-fields associated with randomly organized silver nanoparticles by using electromagnetic calculations based on experimentally fabricated devices, and we clarified the fact that they stemmed from the structural randomness of the silver nanoparticles by adapting a theory of optical near-fields based on an angular spectrum. One of the most interesting challenges in near-field optics and nanophotonics in future will be to gain a deeper understanding of randomness; further detailed modeling, analysis, and even to exploit it and optimize it for practical applications would be expected. We consider that this study paves the way to gaining fundamental insights regarding optical near-field processes associated with randomness in the subwavelength regime.

**Methods**

**Electromagnetic numerical simulations.** Above and beneath the 10 nm-thick silver nanoparticles model area, we assume 100 nm-thick and 90 nm-thick spaces, as schematically shown in Fig. 2a. Continuous-wave (CW) *x*-polarized light, irradiating the surface of the sample from a distance of 92.5 nm, is normally incident on the surface of the silver nanostructures. Absorbing boundary conditions are assumed for the *z*-direction, and periodic boundary conditions are assumed in both the *x*- and *y*-directions. Some of the silver nanoparticles near the boundaries touch the periodic boundaries, meaning that they may be interconnected between upper and lower ends and/or between left and right ends in the computational model. Such conditions may trigger artifacts in the simulations, but we consider that our computational area is large enough, and the total number of nanoparticles in the region is large enough ($N = 468$ particles), and thus the issue of the periodic boundary is marginal. In addition, it is likely that the size of nanoparticles located at the boundaries is smaller, and the periodic boundary conditions are even reasonable in our particular analysis.

**Acknowledgements**

This work was supported in part by Grants-in-Aid for Scientific Research (2330031) and the Core-to-Core Program A. Advanced Research Networks from the Japan Society for the Promotion of Science.

**Author contributions**

M.Naruse, M.O., and M.Naya directed the project; T.T., H.Y., and M.Naya designed experimental devices, and their experimental characterizations.; T.T. and H.Y. performed numerical simulations.; M.Naruse and N.T. conducted theoretical modeling and analysis; M.N. conducted analysis of numerical simulation results.; M.N., T.T., and H.Y. wrote the paper.

**Competing financial interests**

The authors declare no competing financial interests.

**References**

1. Kiyoto, N., Hakuta, S., Tani, T., Naya, M. & Kamada, K. Development of a Near-infrared Reflective Film Using Disk-shaped Silver Nanoparticles. *Fujifilm Res. & Dev.* **58-2013**, 55-58 (2013).

2. Tani, T., Hakuta, S., Kiyoto, N. & Naya, M. Transparent near-infrared reflector metasurface with randomly dispersed silver nanodisks. *Opt. Express* **22**, 9262- 9270 (2014).

3. Granqvist, C.G. Smart Optics. *Advances in Science and Technology (Volume 55)*, Vincenzini P. & Righini G. (eds.) 205-212 (Trans. Tech. Publishing, Switzerland, 2008).

4. Wang, H., Wu, H., Ding, Y. & Zhou, X. Modeling and Analysis on the Cooling Energy Efficiency of Sun-Shading of External Windows in Hot Summer and Warm Winter Zone. *Int'l Conf. Mater. Renew. Energ. & Environ. (ICMREE).* 1086-1090 (2011).

5. Anderson, P.W. Absence of diffusion in certain random lattices. *Phys. Rev.* **109**, 1492-1505 (1958).

6. Mascheck, M., *et al.* Observing the localization of light in space and time by ultrafast second-harmonic microscopy. *Nat. Photonics* **6**, 293-298 (2012).

7. Grésillon, S. *et al.* Experimental observation of localized optical excitations in random metal-dielectric films. *Phys. Rev. Lett.* **82**, 4520-4523 (1999).

8. Ducourtieux, S. *et al.* Near-field optical studies of semicontinuous metal films. *Phys. Rev. B* **64**, 165403 (2001).

9. Birowosuto, M.D., Skipetrov, S.E., Vos, W.L. & Mosk, A.P. Observation of spatial fluctuations of the local density of states in random photonic media. *Phys. Rev. Lett.* **105**, 013904 (2010).

10. Krachmalnicoff, V., Castanié, E., De Wilde, Y. & Carminati, R. Fluctuations of the local density of states probe localized surface plasmons on disordered metal films. *Phys. Rev. Lett.* **105**, 183901 (2010).

11. Le Ru, E.C., Etchegoin, P.G. & Meyer, M. Enhancement factor distribution around a single surface-enhanced Raman scattering hot spot and its relation to single molecule detection. *J. Chem. Phys.* **125**, 204701 (2006).

12. Sapienza, R. *et al.* Long-tail statistics of the Purcell factor in disordered media driven by near-field interactions. *Phys. Rev. Lett.* **106**, 163902 (2011).

13. Homola, J., Yee, S. S. & Gauglitz, G. Surface plasmon resonance sensors: review. *Sensor Actuat. B.* **54**, 3-15 (1999).

14. Jin, R. *et al.* Controlling anisotropic nanoparticle growth through plasmon excitation. *Nature* **425**, 487-490 (2003).

15. Wolf E. & Nieto-Vesperinas, M. Analyticity of the angular spectrum amplitude of scattered fields and some of its consequences. *J. Opt. Soc. Am. A* **2**, 886-889 (1985).

16. Inoue T. & Hori, H. Quantum theory of radiation in optical near field based on quantization of evanescent electromagnetic waves using detector mode. Ohtsu, M. (ed.) 127-199 *Progress in Nano-Electro-Optics IV* (Springer, Berlin, 2005)

17. Naruse, M., Inoue, T. & Hori, H. Analysis and synthesis of hierarchy in optical near-field interactions at the nanoscale based on angular spectrum. *Jpn. J. Appl. Phys.* **46**, 6095-6103 (2007).

18. Naruse, M. *et al.* Optical near-field–mediated polarization asymmetry induced by two-layer nanostructures. *Opt. Express* **21**, 21857-21870 (2013).

**Figure captions**

**Figure 1 | Overview of the fabricated near-infrared light reflection film composed of silver nanoparticles (NASIP (Nano Silver Pavement)). (a)** A picture of the fabricated and deployed near-infrared light reflection film. **(b)** Its thickness is 10 nm, and the average diameter of the silver nanoparticles is about 100 nm to 120 nm. **(c)** Experimentally measured transmittance and reflectance spectra of the device.

**Figure 2 | Electromagnetic computational evaluations based on experimentally fabricated silver nanoparticles. (a)** A schematic illustration of numerical evaluation of silver nanoparticles experimentally fabricated as shown in Fig. 1(b). **(b)** Calculated electric field intensity distributions at a position 5 nm away from the silver nanostructure when the wavelength of the normally incident light was 300 nm to 2000 nm in 100 nm intervals.

**Figure 3 | Dipole-based modeling for randomly distributed silver nanoparticles. (a)** Induced charge distributions in the silver nanostructures, showing magnified view below. **(b)** A schematic diagram of a silver nanoparticle, identified by the index *i*. The position of the center of gravity is given by $G^{(i)}$, and a dipole $\boldsymbol{d}^{(i)} = d^{(i)}\exp(i\phi^{(i)})$ based on the imbalance of induced electron charges with respect to $G^{(i)}$. **(c)** A schematic diagram of the induced dipoles, showing magnified view below.

**Figure 4 | Statistical properties inherent in the induced dipoles. (a,b)** Incidence patterns of **(a)** the magnitude and **(b)** the phase of the induced electric dipoles as a function of the

wavelength of irradiated light. **(c,d)** The average and the standard deviation of (a) the magnitude and (b) the phase are evaluated at each wavelength.

**Figure 5 | Statistical properties of optical near-field distributions associated with the random silver nanoparticles.** The incidence patterns of electric field intensity at planes away from the surface of the silver nanostructure by distances 5, 20, 50, 100, 200, 500, and 1000 nm for each wavelength.

**Figure 6 | Theoretical modeling of optical near-fields based on angular spectrum representation.** **(a)** A schematic diagram of a wave vector and polarization vectors for the angular spectrum representation of optical near-fields. For evanescent components, $\alpha$ takes an imaginary number. **(b)** Orientation of a dipole $d^{(i)}$ and the point of evaluation $r$. **(c)** A two-dipole system located in close proximity. The magnitude and the orientation of each dipole is specified by randomly generated numbers.

**Figure 7 | Characterization of randomness in optical near-fields by the angular spectrum-based theoretical analysis.** **(a,b)** A differently-specified dipole pair gives a different angular spectrum. **(a)** Moderate randomness yields significantly different spectra, whereas **(b)** less randomness gives nearly identical spectra. **(c)** The incidence patterns of the near-field electric field intensity, given by the integral of the angular spectrum, based on dipole pairs that follow random statistics show some variation even as the distance from the silver nanostructures increases. **(d)** On the other hand, with less-random dipole

pairs, such a tendency is not observed. These are clear manifestations of the hierarchical attributes of optical near-fields.

**Figure 8 | Comparison of calculated spectra for the random structure and an ordered one.** Comparison of calculated transmittance and reflectance spectra for the random structure based on the experimentally fabricated silver nanostructure and an "ordered" one, which is virtually constructed in such a way that the sizes of the nanoparticles are the same and the layouts are uniform, whereas the total area and the total number of particles is equivalent to the case of the former random structures. The average reflectance and the average transmittance of the random structure in the wavelength regime between 1000 nm and 2000 nm are respectively higher and lower compared with those of the ordered one.

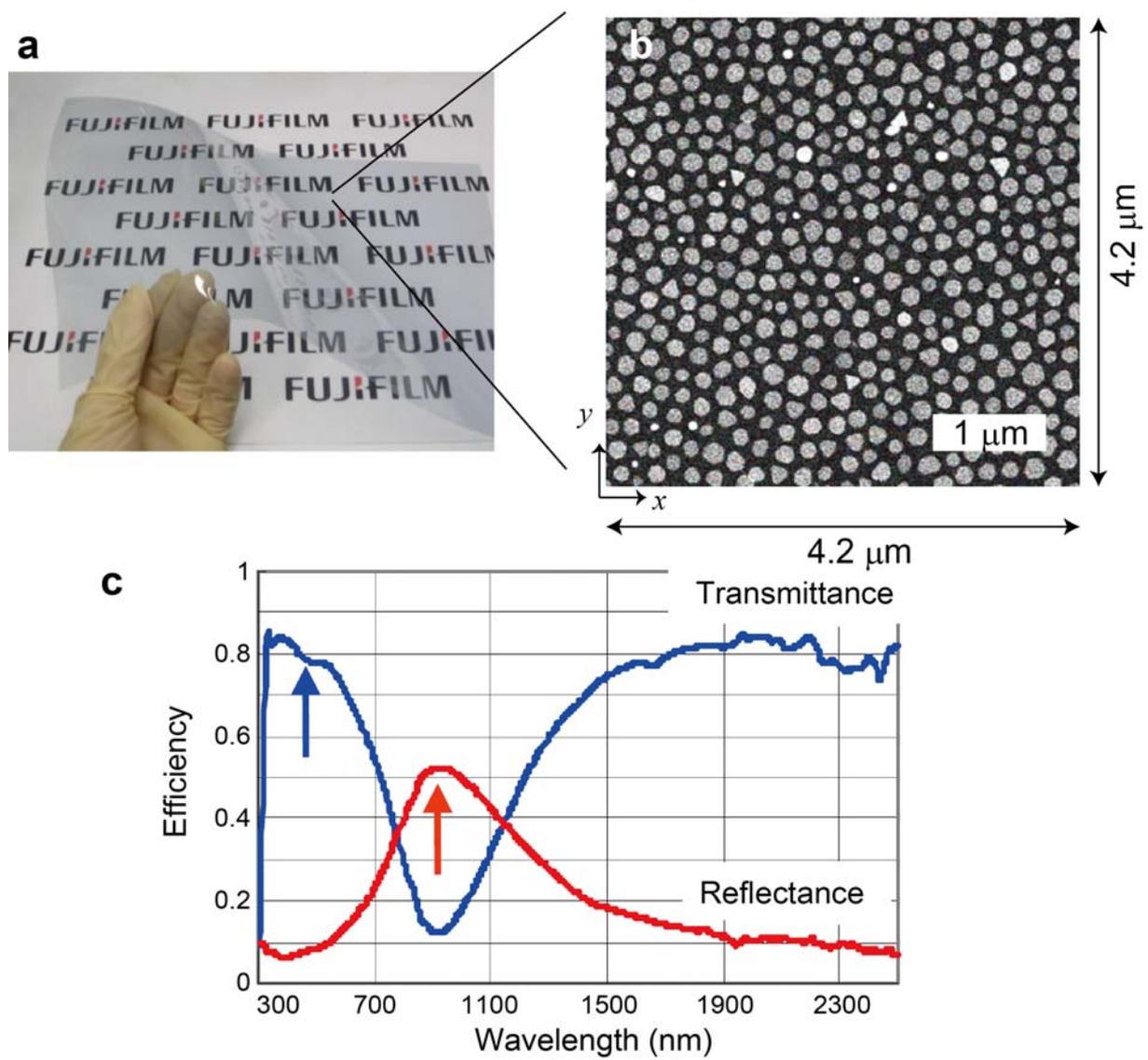

Figure 1

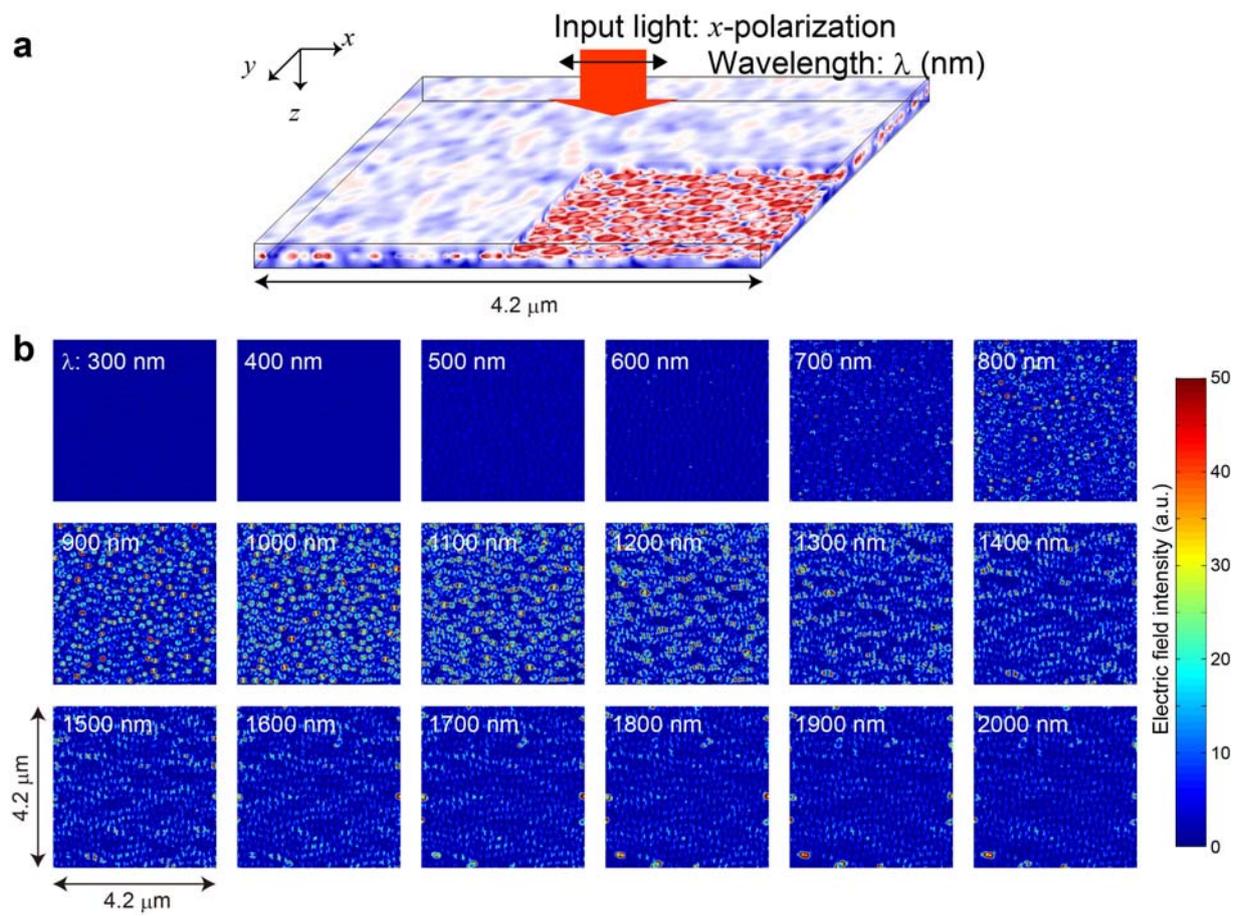

Figure 2

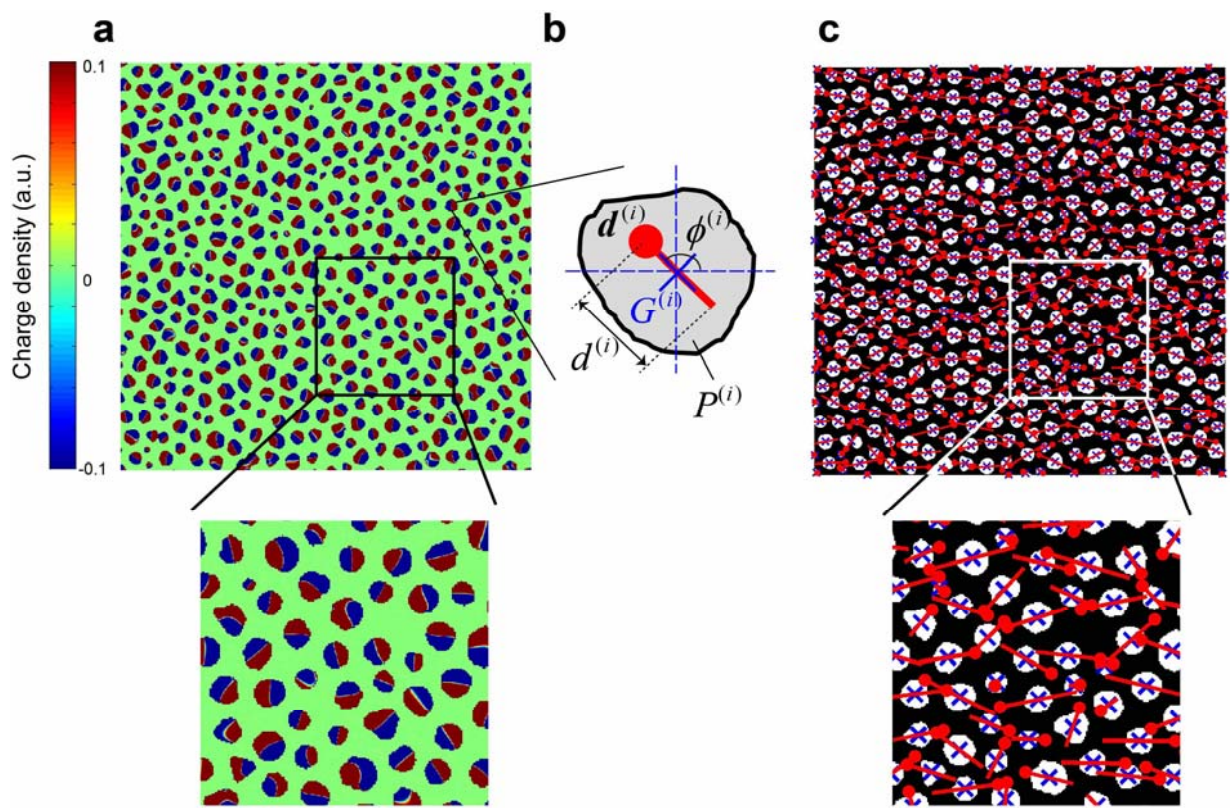

Figure 3

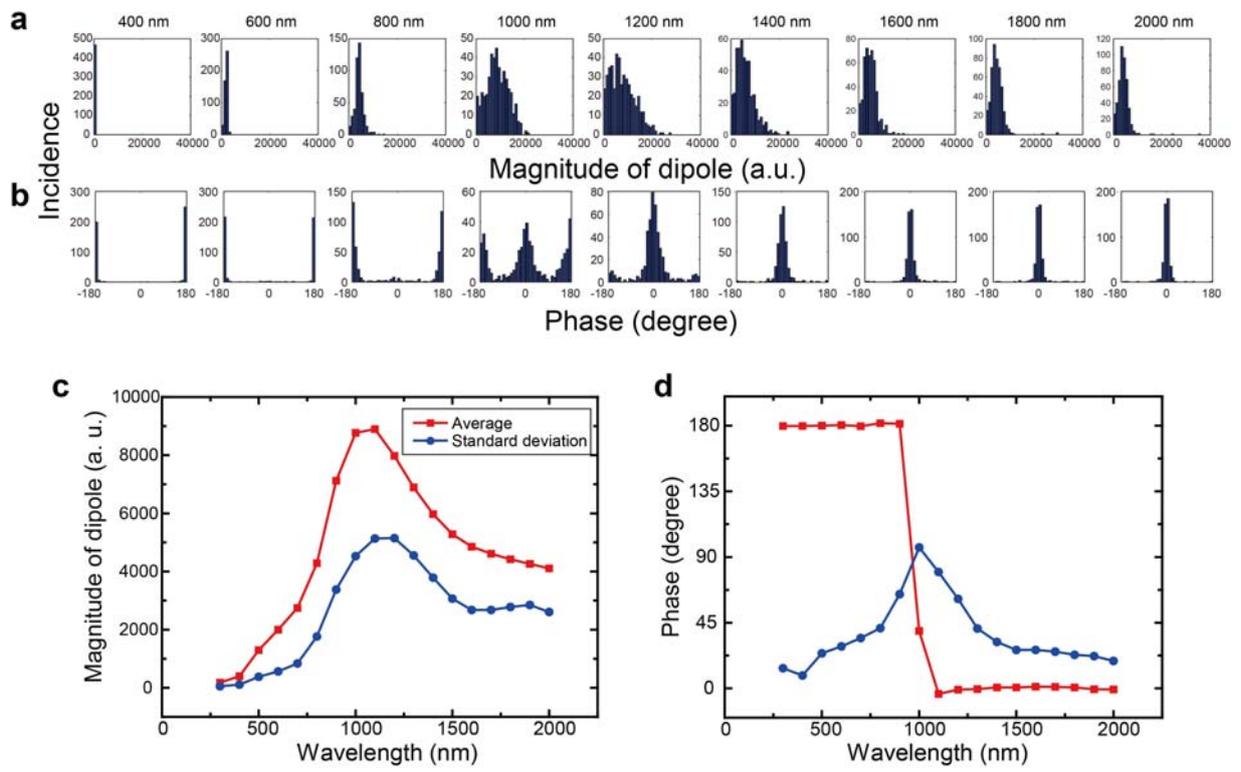

Figure 4

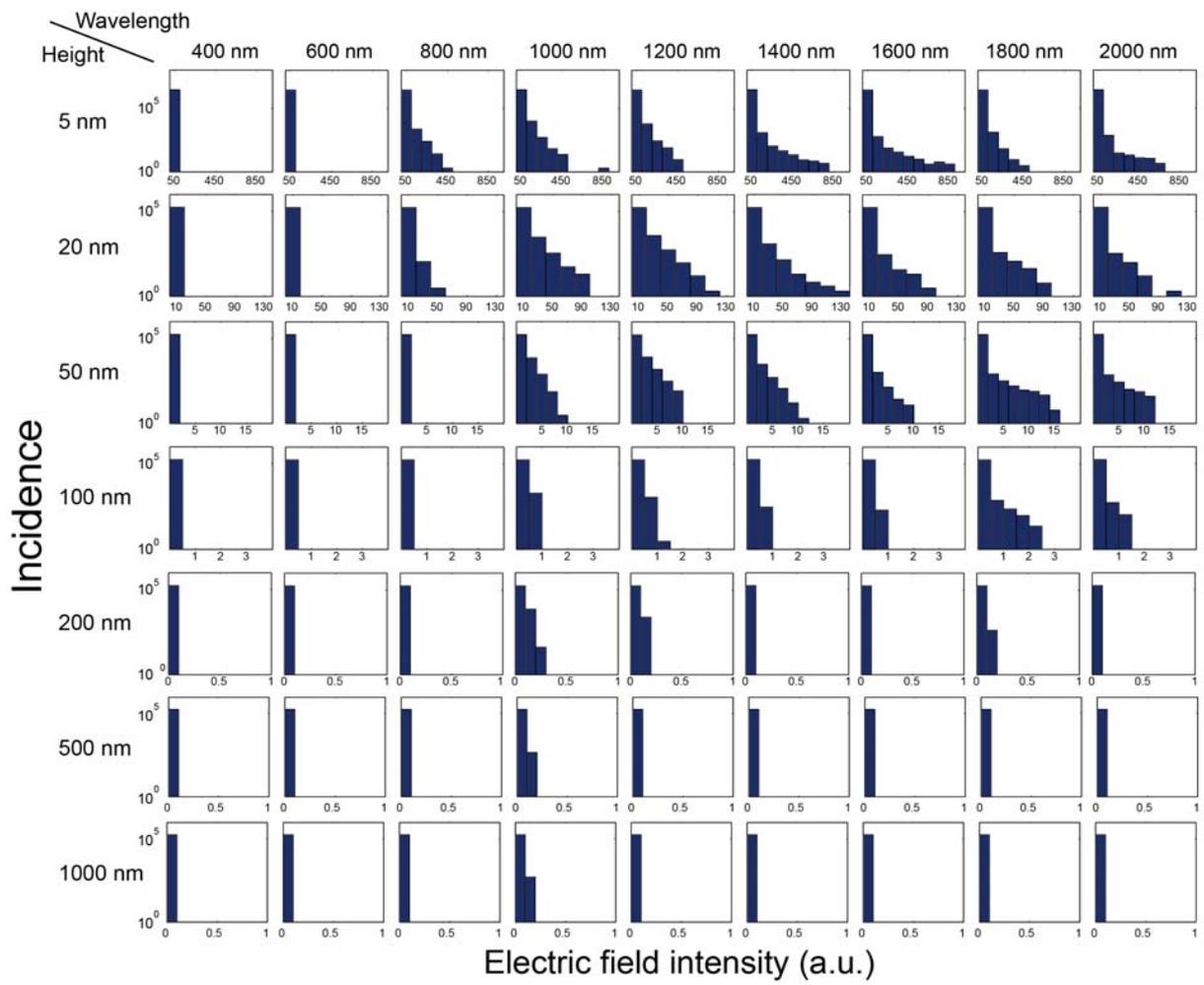

Figure 5

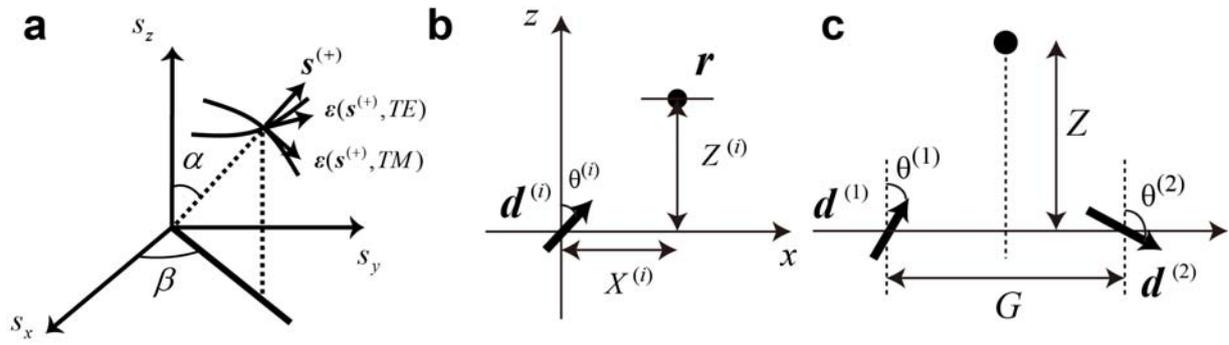

Figure 6

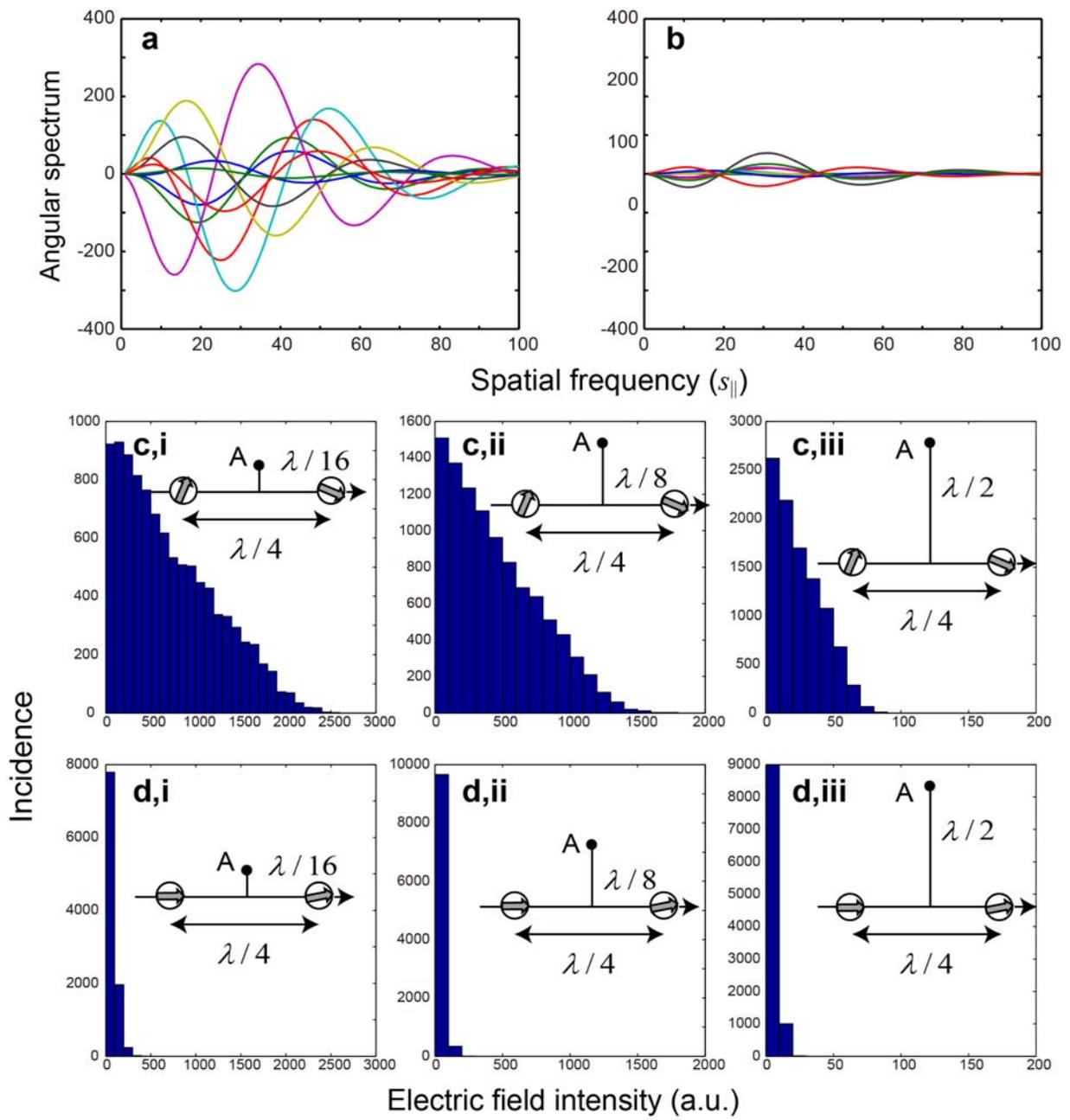

Figure 7

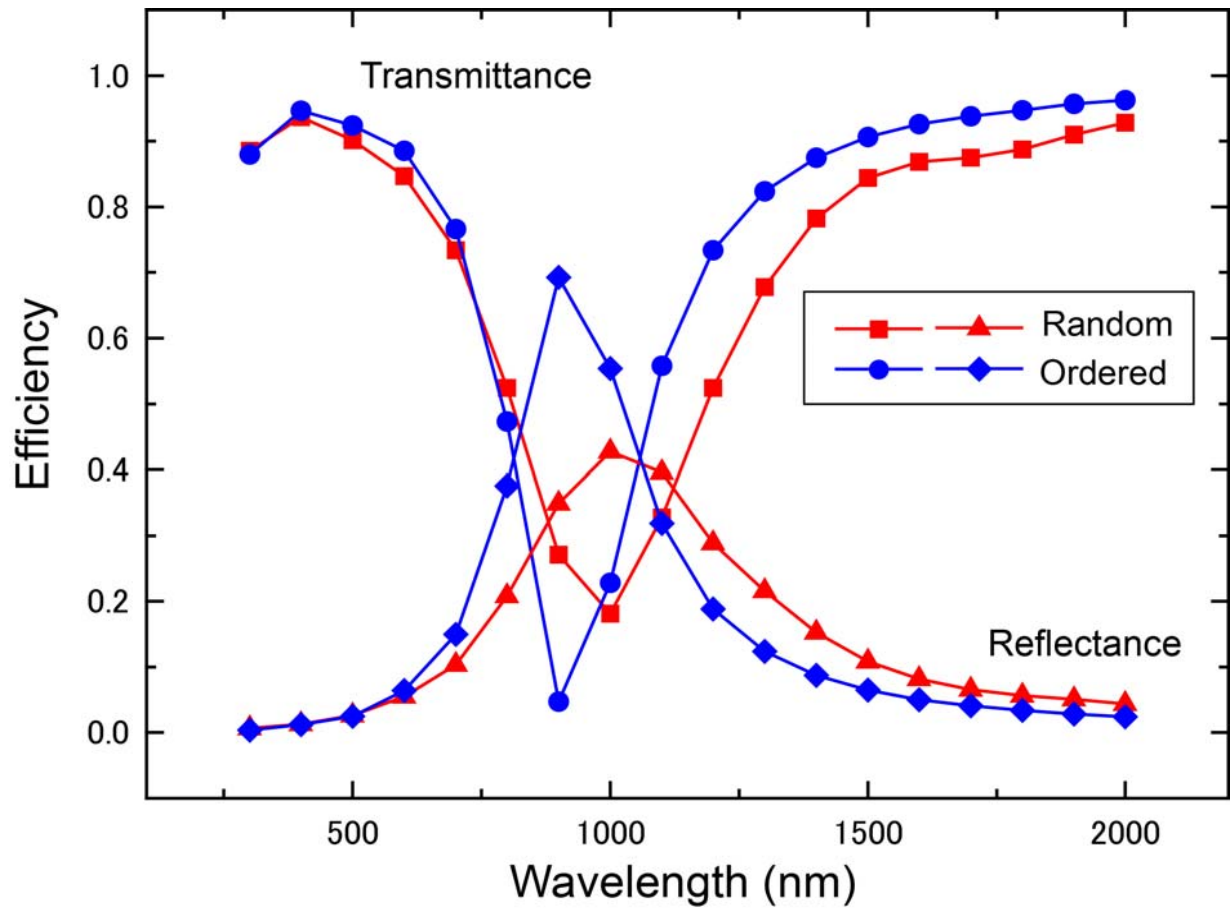

Figure 8